\documentclass[english,aps, prb, twocolumn, superscriptaddress]{revtex4-1}
\usepackage[T1]{fontenc}
\usepackage[latin9]{inputenc}
\usepackage[a4paper]{geometry}
\usepackage{textcomp}
\usepackage{amsmath}
\usepackage{amssymb}
\usepackage{graphicx}
\usepackage{wasysym}
\usepackage{xcolor}
\makeatletter

\providecommand{\tabularnewline}{\\}

\usepackage{hyperref}
\usepackage{amsmath}
\usepackage{amsfonts}
\usepackage{graphicx}
\usepackage{epstopdf}
\usepackage{physics}

\newcommand{\si}[0]{$^{29}$Si}
\makeatother

\usepackage{babel}
\begin{document}
\title{Dynamic nuclear polarization and ESR hole burning in As doped silicon}
\author{J. J\"{a}rvinen} 
\email{jaanja@utu.fi} 
\affiliation{Wihuri Physical Laboratory, Department of Physics and Astronomy, University of Turku, 20014 Turku, Finland} 
\author{D. Zvezdov} 
\affiliation{Wihuri Physical Laboratory, Department of Physics and Astronomy, University of Turku, 20014 Turku, Finland} 
\affiliation{Institute of Physics, Kazan Federal University, Russia}
\author{J. Ahokas}
\author{S. Sheludyakov}
\author{L. Lehtonen} 
\author{S. Vasiliev} 
\affiliation{Wihuri Physical Laboratory, Department of Physics and Astronomy, University of Turku, 20014 Turku, Finland} 
\author{L. Vlasenko} 
\affiliation{Ioffe Institute, Russian Academy of Sciences, 194021 St. Petersburg, Russia}
\author{Y. Ishikawa}
\author{Y. Fujii}
\affiliation{Research Center for Development of Far-Infrared Region, University of Fukui, 3-9-1 Bunkyo, Fukui 910-8507}
\begin{abstract}
\noindent We present an experimental study of the Dynamic Nuclear Polarization (DNP) of \si{} nuclei in silicon crystals of natural abundance doped with As in the temperature range 0.1-1 K and in strong magnetic field of 4.6 T. This ensures very high degree of electron spin polarization, extremely slow nuclear relaxation and optimal conditions for realization of  Overhauser and resolved solid effects. We found that the solid effect DNP leads to an appearance of a pattern of holes and peaks in the ESR line, separated by the super-hyperfine interaction between the donor electron and \si{} nuclei closest to the donor. On the contrary, the Overhauser effect DNP mainly affects the remote \si{} nuclei having the weakest interaction with the donor electron. This leads to an appearance of a very narrow ($\approx$ 3 mG wide) hole in the ESR line. We studied relaxation of the holes after burning, which is caused by the nuclear spin diffusion. Analyzing the spin diffusion data with a simple one-dimensional spectral diffusion model leads to a value of the spectral diffusion coefficient $D=8(3)\times 10^{-3}$ mG$^2$/s. Our data indicate that the spin diffusion is not completely prevented even in the frozen core near the donors. The emergence of the narrow hole after the Overhauser DNP may be explained by a partial "softening" of the frozen core caused by Rabi oscillations of the electron spin.
\end{abstract}
\maketitle

\section{Introduction}

Dynamic nuclear polarization (DNP) is an important technique for achieving
highly polarized nuclear spin states. It has a wide range
of applications in electron spin resonance (ESR), nuclear magnetic
resonance, quantum optics and quantum information processing. Doped silicon is widely considered as a promising system for construction of 
solid state quantum computer \citep{Kane1998,Morton2011,Wolfowicz2015} with the qubits based on nuclear spins of the donor or the surrounding $^{29}$Si nuclei of the host lattice. Finding a method of fast and flexible control on the nuclear spins is what the DNP may offer for the above mentioned applications. Transversal relaxation time of the nuclear spins defines their coherence time and is one of the most important parameters for quantum processing. In the silicon of natural abundance (4.7 $\%$ of \si{}) random flip-flops of the \si{} spins lead to a fast loss of coherence and rapid spin diffusion. Situation is changed near the donors, where the superhyperfine interaction (SHI) with the donor electron creates relatively large shifts of the energy levels of the \si{} nuclei and forms protecting frozen core
around the donor. The coherence times of \si{} inside the frozen
core are about 200 times longer than in bulk silicon \citep{Wolfowicz2015} and may exceed a fraction of a second, long enough for the quantum processing with these spins.
The natural coupling of \si{} with the donor electron inside the
frozen core could be utilized in quantum registers containing several
\si{} nuclei and the donor electron. Further improvement in coherence
times could be attained by increasing the spin polarization and thus
reducing the spin flip-flops around the donor.\cite{Gong2011}

Best conditions for the polarization and manipulation of nuclear spins are realized at low temperatures and in high magnetic fields. This ensures high degree of polarization of electron spins, extremely slow nuclear relaxation, and allows performing  Overhauser (OE) and resolved solid effect (SE) DNP. In our previous work \cite{Jarvinen2014, Jarvinen2015, Jarvinen2017} we realized these conditions at temperatures $<$ 1 K and magnetic field of 4.6 T. We demonstrated efficient control of the donor nuclear spins of $^{31}$P and $^{75}$As using OE DNP. Highly polarized nuclear state was created with fairly low ESR pumping power of $\lesssim$1 $\mu$W. In addition, new and interesting results were obtained in the DNP experiments involving the system of donor electron and \si{} nuclei nearby. We found that the SE leads to an efficient \si{} spin polarization of the closest lattice sites having strongest SHI. 

In this work we conducted a detailed study of the dynamics of the SE and OE DNP of the \si{} nuclei located near enough the donors, called as "interaction zone", to be visible in the ESR spectrum. The OE DNP of these \si{} nuclei is performed by classical hole burning experiment by exciting the inhomogeneously broadened line at some fixed position. We observed very narrow holes with the width of $\approx 3$ mG matching the 2/$T_2$ transversal relaxation rate obtained earlier from the pulsed ESR measurements.\cite{Chiba1972} The observed shape of the holes indicate that the nuclei closest to the donor are not influenced by the OE DNP process. Relaxation dynamics of the holes are analyzed with a simple one-dimensional spectral diffusion model, which provides data on the spectral diffusion coefficient. Owing to a strong magnetic field we are able to make well resolved SE DNP, which leads to an emergence of a pattern of holes and peaks on the original  line shape. The difference in the patterns obtained with SE, based on saturation of the forbidden flip-flip and flip-flop transitions, are used to measure the degree of nuclear polarization of \si{} closest to the donors. We demonstrate that these nuclei can be efficiently polarized using the SE for the whole ESR line covered by the field or frequency modulation. We analyse the observed features and differences of the OE and SE DNP of \si{}, and propose several possible explanations for their mechanisms. We think that the results of this work may be useful for optimization of the methods of the nuclear spin manipulation and help to understand the underlying physics.

\section{Background: DNP in silicon}

DNP is a process where the polarization of electrons is transferred
to the interacting nuclei by the means of electron spin resonance. At the conditions of our experiments, strong magnetic field and relatively low donor densities, we consider two major DNP methods: the Overhauser\citep{Overhauser1953} and the solid effects.\cite{Abragam1985} In the first method the allowed electronic transitions (ESR lines) are saturated, followed by the relaxation via the cross- or forbidden transitions with the change of the nuclear spin. In the SE DNP the flip-flop or flip-flip transitions are excited, which leads to a change of both electron and nuclear spin projections. The details of these two processes are found in numerous detailed texts written about DNP, e.g. refs.\cite{Maly2008,Ramanathan2008,Hu2011,Can2014,Can2015,LillyThankamony2017} In the following we will provide a short summary of the specific features of the OE and SE DNP in doped silicon at high fields and low temperatures. 

We consider a simple model including single electron $S$, a donor
nuclei $I_{d}$ and and single \si{} nucleus $I_{k}$. The index
$k$ indicates the lattice site where the \si{} atom is located.
The Hamiltonian for this system is
\begin{multline}
H=g_{e}\mu_{B}B_0\cdot S-g_{d}\mu_{n}B_0\cdot I_{d}+a_{0}S\cdot I_{d}\\
+g_{n}\mu_{n}B_0\cdot I_{k}+a_{k}S\cdot I_{k}+S\cdot D_{k}\cdot I_{k}.\label{eq:Hamiltonian}
\end{multline}
Constant $a_{0}$ is the Fermi contact interaction energy of donor
nuclei, $a_{k}$ is isotropic or contact superhyperfine interaction
(SHI) energy and $D_{k}$ is dipolar or anisotropic SHI (nonsecular
component) energy tensor for lattice site $k$\cite{Hale1969}
\begin{equation}
(D_{k}){}_{i,j}=g_{e}g_{n}\mu_{B}\mu_{n}\bra\psi\frac{3x_{i}x_{j}-r^{2}\delta_{ij}}{r^{5}}\ket\psi.\label{eq:Dk anisotropic}
\end{equation}

\begin{figure}
\includegraphics[width=0.8\columnwidth]{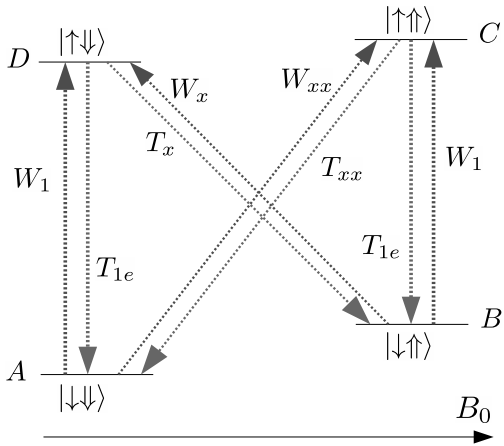}\caption{Simplified energy level diagram of ESR excitation and relaxation processes for electron (single arrows) and \si{} nucleus (double arrows).  $T_{1e}$ is the thermal relaxation time of the electron spin (single quantum) and $T_{x}$ and $T_{xx}$ are the flip-flop and flip-flip (zero and double quantum) thermal cross relaxation times. $W_{1}$  is the allowed ESR transition rate and $W_{x}$ and $W_{xx}$ are the forbidden flip-flop and flip-flip ESR transition rates. The $B_{0}$ axis indicates the relative locations of the
lines in ESR spectrum. \label{fig:DNP-processes-diagram}
}
\end{figure}

The solution of eq. (\ref{eq:Hamiltonian}) without \si{} contribution for $^{75}$As $(I=3/2)$ nucleus
and electron provides 8 energy levels characterized by different projections of the electron and donor nuclear spins. Due to a large spread of the electron cloud of the donor, there are about $1500$ silicon nuclei inside the cloud. For silicon of natural isotope composition (4.7$\%$ of \si{}) there are about 70 \si{} nuclei with spin $(I=1/2)$ having site-dependent SHI with the donor electron. Each of the \si{} splits the donor energy levels to two sub-levels, and the total effect of all the \si{} is an inhomogeneous broadening of the ESR lines, which are composed of individual spin packets with different configurations of the \si{} environment inside its electron cloud. In this work we focus on the DNP of the \si{} nuclei effectively interacting with the donor electron. The own nucleus of the donor is considered only as a spectator which does not participate in the DNP process of the \si{} spin system.
The main spin flip transitions involved in the DNP of the \si{} nucleus
are shown in the simplified energy diagram of fig. \ref{fig:DNP-processes-diagram}, where for simplicity we consider effect of one \si{} interacting with the donor electron.

The anisotropic part of the SHI given by the components of the $D_{k}$, is usually much smaller than $a_{k}$ but it mixes the
wavefunctions with opposite electron spin projection (not shown in
fig. \ref{fig:DNP-processes-diagram}). The order of the mixing angle
can be estimated as\cite{LillyThankamony2017}:
\begin{equation}
\beta_{k}^{2}\approx\frac{(D_{k})_{1,3}^{2}+(D_{k})_{2,3}^{2}}{(4g_{n}\mu_{n}B_0)^{2}},\label{eq:mixing-angle}
\end{equation}
The value of $|\beta_{k}|$ is practically independent on $a_{k}$
if $a_{k}\ll g_{n}\mu_{n}B_0$. Every lattice site for \si{} nuclei
can be assigned to a shell which belongs to one of the 4 symmetry
groups.\cite{Feher1959} All of the nuclei in the shell have the
same $a_{k}$ but the values of $\beta_{k}$ are generally different
and they depend on the direction of the crystal in the $B_{0}$ field.
Values of $\beta_{k}$ for different sites in a shell are calculated
by rotating $D_{k}$ according to the symmetry group of the shell.\cite{Hale1969,Jarvinen2015}

In the OE DNP one of the electronic spin transitions $W_{1}$ is saturated
by the microwave field. If there is a non zero probability for thermal cross
relaxation $T_{x}$ or $T_{xx}$, the nuclear spin population is transferred to state A or B (fig. \ref{fig:DNP-processes-diagram}). Usually the
$T_{x}$ relaxation time (flip-flop or zero quantum) is much shorter
than $T_{xx}$ (flip-flip or double quantum)\cite{Pines1957} which leads to nuclear
polarization even when both of the transitions $A-D$ and $B-C$ are
excited simultaneously. The flip-flop or flip-flip cross relaxations are induced
by the fluctuations of the SHI due to molecular
motion in liquid samples, electrons in conducting solids or lattice phonons in insulators.\cite{Can2015,Overhauser1953, LillyThankamony2017,Pylaeva2017}

The SE DNP is performed by exciting the forbidden ESR transition $W_{x}$ or $W_{xx}$ followed by the much faster thermal relaxation of the electron spin with the relaxation time $T_{1e}$. The SE polarization rates are set by the forbidden transition rates $W_{x}^{(k)}$ and $W_{xx}^{(k)}$ which are dependent
on the value of $\beta_{k}$ and microwave magnetic field $B_1$ according to Fermi's Golden Rule 
 
\[
W_{x}^{(k)}=W_{xx}^{(k)}\propto B_{1}^{2}|S_{x}|^{2}\propto B_{1}^{2}\beta_{k}^{2}.
\]

The high static $B_{0}$ field reduces the efficiency of SE DNP by
reducing $\beta_{k}$ as shown in eq. (\ref{eq:mixing-angle}). The
high $B_{0}$, on the other hand, is needed for the resolved SE DNP where the SE transitions are outside of the main ESR lines. For natural silicon this requires a field larger than $\hbar\omega_{1/2}/g_{n}\mu_{n}\approx1.3$ T, assuming
ESR linewidth $\omega_{1/2}$ of about 3.6 G. Other parameter which defines the SE DNP efficiency is the
microwave $B_{1}$ field. The $B_{1}$ field increases as $\sqrt{P}$,
where $P$ is the ESR excitation power. Operating at temperatures
below 1 K limits the applicable ESR excitation powers to few
$\mu$W and makes it difficult to achieve high SE DNP rates.

Both the OE and SE DNP modify the \si{} spin polarization around the donor, which is seen in the ESR lineshape as a pattern of holes and peaks, dependent on the actual pumping procedure. First DNP experiments on silicon were reported by Feher and Gere \cite{Feher1959a}
in 1959. The authors burned a hole on ESR line of As doped silicon.
Due to the low magnetic field of 0.3 T holes for both OE and SE DNP
were created at the same time. This observation was explained correctly
only after several years.\cite{Marko1970} Holes caused by polarization
of \si{} were later studied in Boron-doped Si with X-band ESR and temperatures
above 1 K.\cite{Dirksen1989} In this work satellite peaks around
the burned holes were observed and they were attributed to the SE polarization of \si{}.

Nuclear relaxation and spin-diffusion are the processes which return nuclear polarization to the equilibrium after its modifications by DNP. The relaxation process is usually limited by the nuclear Orbach process, which has strong exponential dependence on temperature, and below 1 K this effect is negligibly slow.\cite{Abragam1985} The spin diffusion, introduced by Bloembergen\cite{Bloembergen1949} in 1949, occurs due to the mutual flip-flops of the spins of the neighbouring nuclei. It leads to a propagation of the \si{} nuclear polarization into the bulk of the host crystal. The rate of spin diffusion is relatively high far away from the donors. It slows down upon approaching the donors due to the increase of the SHI energy difference of the \si{} spins pairs, and is completely forbidden inside the frozen core near the donors where the difference is largest. The radius of the frozen core is $\approx$8 nm, which is given by the distance at which the difference in the SHI energy of the nuclei is equal to the width of the NMR line.\cite{DeGennes1958,Khutsishvili1966}

\section{Experiments}\label{exp}

The sample studied in this work is a $6\times6\times0.4$
mm natural silicon crystal with (001) crystal axis parallel to magnetic
field, doped with $^{75}$As at concentration $n=3\times10^{16}$ cm$^{-3}$. The sample is mounted on a flat mirror of semi-confocal Fabry-Perot resonator which is thermally linked to a mixing chamber of a dilution refrigerator. Minimum temperature of 300 mK in this work was limited by the heat load through the waveguide connecting the resonator to the ESR spectrometer. The resonator with the sample is placed in the center of high-homogeneity superconductive magnet operating in persistent mode at the field of $B_{0}=$4.6 T. A separate coil is used to sweep magnetic field $\Delta B$ through the resonance lines. We do not use any field modulation due to eddy current heating induced at low temperatures. A cryogenic ESR spectrometer\cite{Vasilyev2004} provides high sensitivity at fairly low ESR power which is important at low temperatures when the lines are easily saturated. Spectrometer has maximum power of $\approx$10 $\mu$W incident into the cavity which generates RF field of $B_{1}\approx 18$ mG. The measured $T_{1e}$ time of the donor electrons is about 0.6
s, and in order to avoid strong saturation effects we have to decrease the incident power by 70-75 dB.  Additional details of the experimental setup can be found in ref.\cite{Jarvinen2017}

\begin{figure}
\includegraphics[width=0.9\columnwidth]{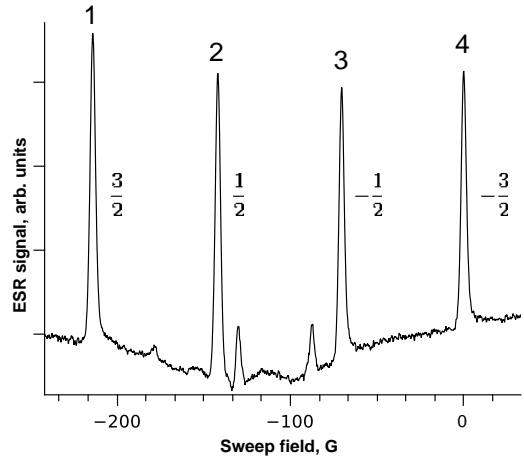}\caption{Si:As ESR spectrum at 300mK. Line numbers are referred in the text
and the fractions indicate the spin state of As nucleus. Two small ESR
lines at the center are due to P impurities.\label{fig:ESR-spectrum}}
\end{figure}

The full quartet of the ESR spectrum of $^{75}$As donors is shown in fig. \ref{fig:ESR-spectrum}. The lines are separated by hyperfine interaction of about 72 G between the donor electron and nuclear spins. The lines are inhomogeneously broadened by the SHI with
\si{} located in the interaction zone, providing the full width at half maximum of $\approx$ 3.6 G. From the width of spectral holes described below we estimate our field/frequency resolution to be about 1 mG/3 kHz. This is partially limited by the spectral width of the mm-wave source used in the ESR spectrometer.\cite{VDI} The other source of the hole broadening occurs at long pumping times and is caused by a small $\sim 5 \mu$G/s drift of the field of the main magnet. The sample also contains P atom impurities. A doublet of their ESR lines is visible in the center of the whole ESR spectrum (fig. \ref{fig:ESR-spectrum}).

Experiments presented in this work typically start with the preparation of the sample in a certain spin state of the donor nucleus using the $^{75}$As DNP. The OE or SE DNP can be efficiently used for creating high degree of nuclear polarization of the donor nuclear spins, which we demonstrated in our previous studies.\cite{Jarvinen2014, Jarvinen2017} We use this technique to transfer spin population to one of the four lowest energy levels of $^{75}$As with certain nuclear spin projection of the donor. This modifies the quartet of the $^{75}$As ESR lines in the following way. Pumping the fourth line (-3/2 spin state) followed by the $^{75}$As OE via the flip-flop transition ($T^{\text{As}}_{x}$) moves the line area to the third. Subsequent pumping of the third line moves it into the second and finally to the first. Moving all lines into the first one, provides a factor of four increase of the signal intensity. At the same time it is beneficial to have all spins in the state  with spin projection 3/2 since the OE DNP of the donor nuclear spins in the reverse direction proceeds via the flip-flip forbidden transition ($T^{\text{As}}_{xx}$), which works much more slowly. As a result, pumping the first line with maximum power does not lead to a substantial transfer of area back to the other lines.\cite{Jarvinen2017} Nuclear spin state remains unchanged in this case, and the \si{} DNP is not mixed with that for the donor. Typically the nuclear spins of donors in the sample were polarized to the 3/2 state in the beginning of the experimental run. Then, only the first line in the ESR spectrum remained, and persisted for several days due to extremely low relaxation time of donor nuclei \cite{Jarvinen2017} at the low temperatures used in this work. 
In order to increase the pumping efficiency for the whole ESR line we applied a frequency modulation to our mm-wave source with the rate of 50-100 Hz and frequency deviation of 15-20 MHz, wide enough to cover whole 3.6 G wide line.

In the following we will describe the ESR hole burning experiments utilizing either OE or SE. The evolution of the holes during the pumping and recovery to equilibrium is sufficiently slow at temperatures below 1 K, which creates ideal conditions for a detailed study of the DNP mechanisms. In addition we attempted to create highly polarized \si{} state of nuclei located inside the electron cloud of the donor. For that, we used the SE based on the excitation of the flip-flip ($W_{xx}$) transition. The sweep of the magnetic field was stopped at the position of this transition, which is located 13.9 G to the low field side of the ESR line center. We applied frequency modulation of the mm-wave source to cover the whole line as mentioned above. Since the forbidden transition probability is rather low in the case of pumping with the modulated source, we had to apply pumping for 2-3 days to get a significant effect. We will demonstrate below that the SE hole burning can also be effectively used to characterize the polarization of the nuclei located in the interaction zone inside the electron cloud of the donor.

\subsection{Solid effect DNP}

\begin{figure}
\includegraphics[width=0.9\columnwidth]{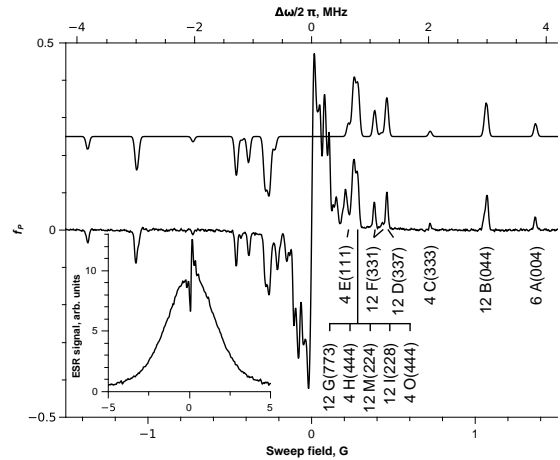}\caption{Solid effect pattern after 300 s ESR excitation at 280 mK. The inset
shows the pattern in wider sweep together with the high field As ESR
line (nr. 4 in fig. \ref{fig:ESR-spectrum}). Calculated spectrum
for inner lattice sites is shown in the upper trace (shifted up by
0.25 units). The number of lattice sites in shells A-O used in the
calculation are marked on the corresponding peaks at lower right together
with the lattice site coordinates.\label{fig:SE-shells}}
\end{figure}

\begin{figure*}
\includegraphics[width=0.99\textwidth]{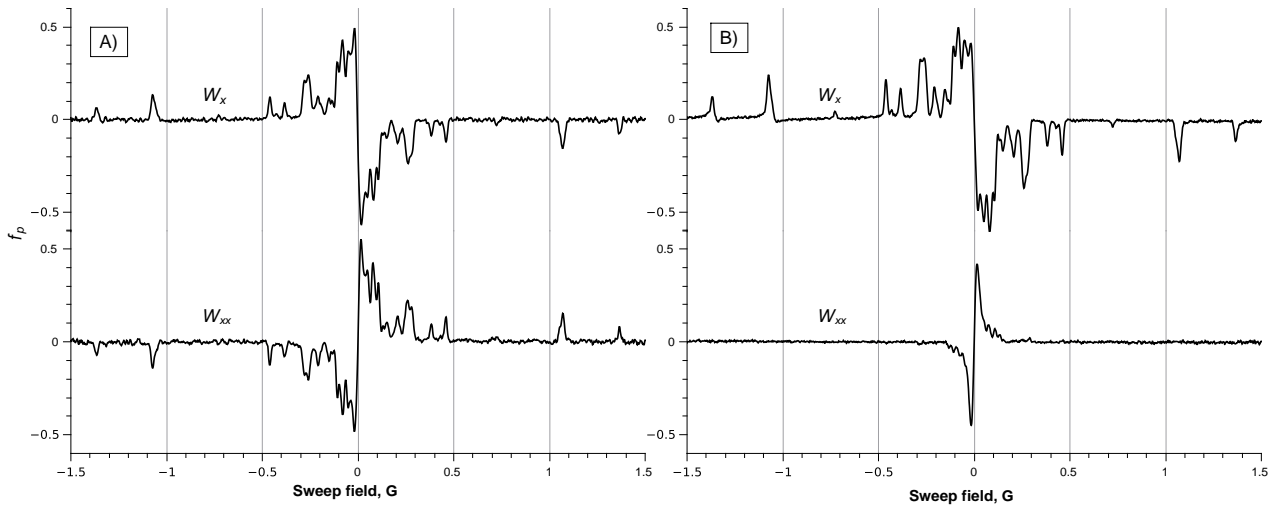}

\caption{SE hole burning for unpolarized A) and polarized B) \si{} nuclei. Upper traces are $W_x$ (flip-flop) and lower traces $W_{xx}$ (flip-flip) SE DNP patterns. The SE burning time was 300 s for each figure.\label{fig: SE-hole-burning}}
\end{figure*}

The SE DNP is performed by partial saturation of $W_{x}$ or $W_{xx}$
transition with simultaneous flips of the electron and nuclear spins. The transitions are located at about $\pm13.9$ G from
the ESR lines in 4.6 T field and they are too weak to be visible
in the ESR spectrum. Since the SE pumping position is located relatively far from the ESR lines, there is no danger to change the populations of the donor nuclear states and therefore the \si{} SE may be done for any ESR line of $^{75}$As. The DNP experiment proceeds first by measuring the undisturbed ESR signal $f(\Delta B)$. The magnetic field is then tuned to $W_x$ or $W_{xx}$ transitions and the full ESR power is used for excitation without frequency modulation. Typically we used pumping times in the range of 10 to 1000 s. After the pumping the line shape $f_{SE}(\Delta B)$, which contains the pattern of holes and peaks created by the SE, is measured. The center of the pattern is located at $\Delta B=0$. The normalized SE pattern $f_{p}(\Delta B)$ is then extracted by subtracting and scaling with the undistorted line shape.

An example of a SE hole-peak pattern is shown in fig. \ref{fig:SE-shells}. The pattern was created by exciting $W_{xx}$ transition at -13.9 G from the ESR line with the maximum available mm-wave power for 300 s. The amplitudes of the holes or peaks did not grow significantly larger for longer pumping times. Similar patterns were observed in our previous work for $^{31}$P donors in silicon. \cite{Jarvinen2015} They were interpreted by the DNP of \si{} nuclei located at certain lattice sites near the donors, and having specific SHI with the donor electron. The redistribution of the spin packets occurs due to the spin flips of the \si{} nuclei and results in a transfer of the ESR line area from the holes into the peaks. The frequencies of the hole and the corresponding antisymmetric peak are given by 
\begin{equation}
\omega_{k}\hbar=\omega_{s}\hbar\pm\frac{a_{k}}{2},\label{eq:hole separation}
\end{equation}
where $\omega_{s}/2\pi$ is the resonance frequency of the center of the pattern and $ a_{k}$
is the SHI energy.

Each peak-hole pair in fig. \ref{fig:SE-shells} corresponds to polarization
of nuclei in a group of symmetrical lattice sites, called as shells, around the donor.\cite{Jarvinen2015}  The shells with large separation are marked from A to O.\cite{Feher1959a} The peaks with separation smaller than about 300 mG cannot be resolved from each other. In SE DNP the flipping probability of a nucleus depends on the dipolar (anisotropic) SHI which can be roughly estimated from eq. (\ref{eq:mixing-angle}). The values of $D_{k}$ for the nuclei in the same shell are slightly different and depend on the orientation of the crystal in the magnetic field. Therefore, each component in the SE pattern (hole or peak) could have some structure containing several overlapping or even split features. This is the reason for the broadening of the holes and peaks in the SE pattern. The SE patterns observed in this work, match well to the more accurate calculations \cite{Jarvinen2015} which are presented as a solid line in the upper trace of fig. \ref{fig:SE-shells}. The decay of the SE pattern is barely measurable during observation of several days at temperatures below 500 mK.

Let's now consider the influence of the initial \si{} polarization on the SE patterns. Starting with the sample of unpolarized \si{} implies that the populations of the A and B levels in fig. \ref{fig:DNP-processes-diagram}
are equal, and the polarization $P^{(k)}_0=(N^{(k)}_B-N^{(k)}_A)/(N^{(k)}_B+N^{(k)}_A)=0$ for each lattice position labeled with index $k$. The area of the peak-hole pairs, depends on the initial polarization $P_0^{(k)}$ and, for $W_{xx}$, is proportional to $\sum_{k}(1-P_{0}^{(k)})/2$, where summing is taken over all the lattice sites contributing to the SE peak. Similarly, the strength of the peak-hole pairs pumped via the $W_{x}$ channel will be proportional to the initial polarization as $\sum_{k}(1+P_{0}^{(k)})/2$. In case of zero initial polarization antisymmetric patterns having equal amplitudes should be produced for the $W_x$ and $W_{xx}$ SE. If the initial polarization equals to 1, i.e. all the population is already in the spin state B, then pumping the $W_{xx}$ transition does not produce any effect and no patterns should be visible. Pumping of the $W_{xx}$ transition in this case produces twice larger amplitudes of the peak-hole pairs compared to the unpolarized case.  

This behavior is well confirmed by our experiments presented in fig. \ref{fig: SE-hole-burning} where we present the patterns recorded after the SE DNP via the flip-flop ($W_x$) and the flip-flip ($W_{xx}$) transitions for comparison. The SE patterns presented on the left side (A) are measured in the beginning of the experimental run, immediately after the cool down, having a fresh unpolarized sample. The $W_x$ pattern is antisymmetric to the $W_{xx}$ one and the amplitudes of the peaks and holes are nearly equal, as it should be for an unpolarized sample. The situation changed dramatically after we performed the SE aiming on the polarization of the whole sample. As described in Section \ref{exp}, we pumped the flip-flip transition with modulation covering the whole ESR line for 2-3 days. The SE patterns recorded after this procedure are presented in fig. \ref{fig: SE-hole-burning} B. One can see that the remote features in the flip-flip pattern ($W_{xx}$, lower trace) disappeared, leaving only a dispersion-like structure in the center, while the flip-flop pattern amplitude ($W_x$, upper trace) has nearly doubled. The remaining unresolved hole and peak pattern in the center can be explained by poor polarization of the nuclei with weak SHI with its electron. Thus, the efficiency of the SE DNP is largest for the closest \si{}, with strong anisotropic SHI. The natural shape of the As ESR line is lost after the SE pattern is created and the relaxation of the pattern is extremely slow. The
patterns can be ``cleaned'' from the ESR line by exciting \si{} NMR transitions, which, however, also modifies the \si{} polarization. Due to this reason, the $W_{x}$ and $W_{xx}$ patterns in fig. \ref{fig: SE-hole-burning} were created on different As lines (e.g. 3 and 4).
 
\subsection{Overhauser effect DNP}

\begin{figure}
\includegraphics[width=0.8\columnwidth]{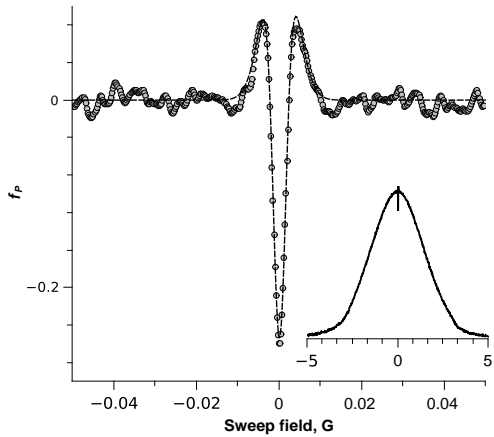}

\caption{OE hole burned for 50 s on ESR line of As doped silicon at 420 mK.
The dashed line is fitted function to the data points. Inset shows
the hole with the main ESR line. \label{fig:OE-hole-burned}}
\end{figure}

\begin{figure}
\includegraphics[width=0.8\columnwidth]{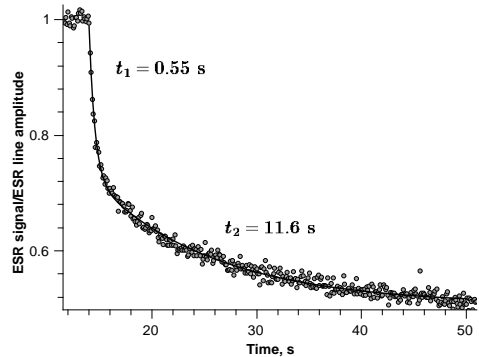}

\caption{ESR signal evolution during OE hole burning. The line is double exponential fit to the data points. The decay times $t_{1}$ and $t_{2}$ are
extracted from the fit.\label{fig:ESR-saturationl-evolution}}

\end{figure}

The OE DNP of the electron-nuclear system of \si{}-As is conducted by saturating the allowed $W_{1}$ transitions, which is equivalent to a classical hole burning experiment. If the \si{} nuclear spin configuration surrounding the donor does not change, one should observe a hole with the width defined by the electron transversal relaxation rate $2/T_2$, and which should vanish with the characteristic time of spin-lattice relaxation ($T_{1e}\approx 0.6$ s). Indeed, a narrow hole is observed as a result of even partial saturation anywhere on the ESR line (fig. \ref{fig:OE-hole-burned}). Like in the SE experiments described above, the hole lived for much longer time than $T_{1e}$ implying flips of the \si{} spins inside the observation zone near the donor. In the example of fig. \ref{fig:OE-hole-burned}, the OE hole was created in the center of the 1st ESR line by stopping the sweep there for 50 s with very low power $\sim 1$ pW used for detection. In fig. \ref{fig:ESR-saturationl-evolution} we present the evolution of the ESR signal during burning. One can see that the partial saturation of the ESR transition occurs at the time scale of a fraction of a second, after which a slow decrease of the signal indicates the growth of the hole. The latter is caused by the transfer of the spin packets due to the flip-flop or flip-flip relaxations of the donor electron and \si{} nuclear spins. However, we did not observe any patterns of peaks on either side of the hole, similar to that of the SE DNP, which can be seen in the inset of fig. \ref{fig:OE-hole-burned}. Instead, two narrow shoulder peaks were detected on both sides of the hole. 

A good empirical fit to the observed pattern of the hole and shoulder peaks
is provided by a sum of two Gaussian functions

\begin{multline}
f(\Delta B)=A \left( \exp(-\ln2\frac{(\Delta B-B_{1})^2}{w_{1}^2})\right.\\
\left.-\frac{w_{1}}{w_{2}} \exp (-\ln2 \frac{(\Delta B-B_{2})^2}{w_{2}^2})\right),\label{eq:Hole-fit_function}
\end{multline}

where $B_{1}$ and $B_{2}$ are the locations and $w_{1}$
and $w_{2}$ are the half widths of the Gaussian functions. An example of the fit is shown in fig. \ref{fig:OE-hole-burned}.

In OE hole burning, any spin flip of \si{} should move the resonant spin packet from the hole to some other place in the ESR line, dependent on the value of the SHI for the flipped spin. The rates of this transfer are proportional to the cross relaxation times $T_{x}$ and $T_{xx}$ (fig. \ref{fig:DNP-processes-diagram}) which are also specific to the actual locations of the flipped \si{} near the donor. One would expect that this should lead
to multiple peaks on both sides of the hole separated by the SHI corresponding to a certain lattice sites. Instead, we observed just a strong narrow hole with two adjacent peaks.
Comparing the area under the hole and the shoulders (fig. \ref{fig:OE-hole-burned}), we found that they are equal within our measurement accuracy of about 10 \%. This shows that the OE DNP leads to the transfer of the spin packets only to the adjacent peaks near the hole, and not to any other locations. The spins with rather weak SHI, located
relatively far from the donor, are the ones which are flipped during
the OE hole burning. 

This result is unexpected since the OE DNP is dependent on the thermal cross-relaxation, which at low temperatures is a result of
thermal modulation of the isotropic or anisotropic SHI by the lattice phonons. \cite{Overhauser1953,Pines1957,Jeffries1960,Castner1967,Can2014,LillyThankamony2017} In this case, the rate of cross-relaxation is largest for the \si{} with strongest SHI, and is given by \cite{Pines1957}

\begin{equation}
T_{x}^{-1}\propto Ta_{k}^{2},\label{eq:The-OE-DNP-rate}
\end{equation}

where $T$ is temperature. As one can see in fig. \ref{fig:OE-hole-burned} the only observable peaks appeared within $\pm10$ mG from
the hole, indicating that spin flips occur mostly in the group of
nuclei having SHI $a_{k}/h\lesssim28$ kHz. For these nuclei the relaxation rate $T_{x}^{-1}$ is about $7\cdot 10^4$ times smaller than for the lattice site (400), which has the strongest SHI of 7.6 MHz. It is peculiar that only the weakly interacting spins are affected by the OE hole burning, and we will further refer to this as Flipping of Weakly Interacting Spins (FWIS) effect.

One may think that the reduction of the cross-relaxation rate for remote \si{} nuclei can be compensated by a larger number of these nuclei which increases as $N\sim r^2$ with the distance from the donor. However, it turns out that this factor does not outweigh the decrease of the SHI. 
An order of magnitude estimate of the OE efficiency as a function of the SHI including both factors can be done using eq. (\ref{eq:The-OE-DNP-rate}). The rate of OE DNP burning at distance $r$ from the donor is $\dot{N}(r)= N(r)/T_x(r)\propto a_{k(r)}^2 T N(r)$. For the donor electron, we can use
hydrogen like wavefunction\cite{Denninger1997} $\psi(r)\propto\exp(-\frac{r}{r_{B}})$,
which gives for isotropic SHI $a_{k}=a_0 |\psi(r_{k})|^{2}$,
where $r_{k}$ is the position of \si{} nucleus and $r_{B}$ is the effective Bohr radius for Si:As ($\approx1$ nm). As the number of \si{} nuclei $dN$ at distance $r$ from the donor is proportional to $r^{2}dr$, the OE polarization rate is
\[
dNT_{x}^{-1}\propto r^{2}T_{x}^{-1}dr=-(\frac{r_{B}}{2})^3(\ln\frac{a_{0}}{a_{k}})^{2}Ta_{k}da_k.
\]
This function has a maximum at $a_k=a_0/e^2\approx1$ MHz, when the highest SHI value is taken for $a_0$. 

In reality, the dependence between $r_{k}$ and $a_{k}$ is not so simple function as assumed here. A more accurate calculation would take into account
that there is a considerable spread in possible $r_{k}$ for different
$a_{k}$ due to strongly oscillating nature of the electronic wavefunction.\cite{Kohn1955} This would, however, only add extra complicity and will not change
the conclusion due to the weak logarithmic dependence of $N$ on $a_k$. 

The minimum observed width of the hole allows us to make an estimate of the maximum distance at which the \si{} spin flip occurs, so that the effect is still visible by the ESR. This distance corresponds to the edge of the observation zone, which is defined as a region around the donor where the flips of the nuclei provide visible changes of the ESR line shape. A rough estimate of the effective radius of the observation zone may be obtained by finding the distance at which the SHI energy equals to the minimum width of the hole. Using the above presented  approximation for the dependence of the $a_k$ on the distance to the donor, we get the characteristic radius of the observation zone $r_o \approx 5$ nm. This value is smaller than the $\approx 8 $ nm radius of the frozen core\cite{Guichard2015}, which implies that all the effects of the spin dynamics leading to emergence of the hole occur inside the frozen core where the spin diffusion is strongly suppressed.

We propose a possible explanation of the FWIS-effect which is related to "softening" of the frozen core due to the Rabi-oscillations of the electron spin when the ESR transition is saturated. The power used for OE hole burning is very small, only about 1 pW. The estimated excitation field during burning is about 3 $\mu$G, which gives electron Rabi frequency of 8 Hz, about four times larger than $1/T_{1e}$. For the \si{} nuclear spins this is seen as an oscillating magnetic field, and if this oscillation period is smaller than the flip-flop time with the nearby nuclei, the SHI is averaged to zero, and the spin diffusion becomes allowed. The frozen core is "softened" for the spin ensemble which is on resonance. Physically the spin flips occurs somewhere at the edge of the observation zone. The nuclei in this region may now exchange their spin state with the neighbors, after which their corresponding spin packets resonance frequencies change out of the resonance, back to the "frozen" spin-diffusion state. Eventually this leads to the appearance of the hole with peak on both sides. Equal amounts of the \si{} spins are flipped and flopped, thus, the average polarization does not change. This mechanism is not related with DNP, but may explain why the spin packets move only close to the pumping point in the ESR line. 

The hole shape observed in this work for $^{75}$As doped silicon is somewhat different from the holes seen in our previous measurement with $^{31}$P, where the shoulder peak on the right side of the hole was larger than the left for the sample with low \si{} polarization.\cite{Jarvinen2015} This was explained as a consequence of the OE DNP with the relaxation via the flip-flop transition ($T_x$ - channel). However, that sample had higher concentration $(\approx 1.5 \times 10^17)$ of donor atoms and rather large cluster peak at the center of the ESR spectrum. This possibly leads to polarization processes, similar to OE, through other possible channels opened by electron-electron interactions of the donors.\cite{Chiba1972}

\begin{figure}
\includegraphics[width=1\columnwidth]{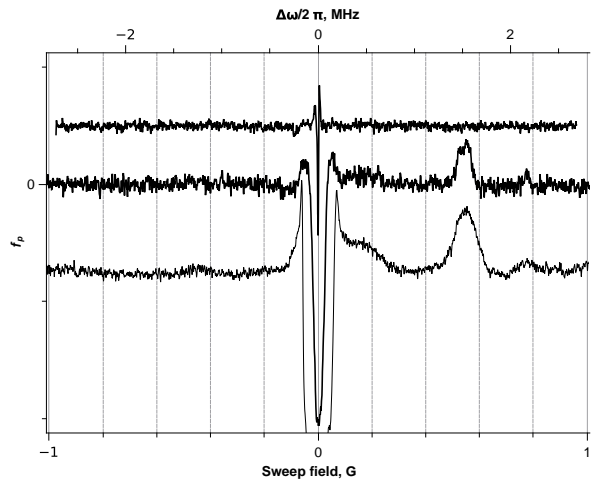}

\caption{OE hole burning on the As ESR line. 1st trace from the top: OE pattern burn time
1000 s at 424 mK. 2nd trace: OE pattern with 30 kHz modulation and
burn time 1800 s at 700 mK. 3rd trace: OE pattern with 100 kHz modulation
and burn time 2000 s at 700 mK. The lines are shifted vertically from each other
for clarity.\label{fig:OE_hole_mod}}
\end{figure}

To get further insight into the OE DNP, we added a small FM modulation with 50 Hz rate to ESR excitation during the OE hole burning. The modulation widens the hole burning region proportionally to the FM modulation amplitude. In this case, even with a fairly small FM deviation of 30 kHz, we observed two peaks on the high field side of the hole (second trace in fig. \ref{fig:OE_hole_mod}). The peaks become stronger and wider with the increasing FM deviation. The strongest peak is shifted to higher fields by $\approx0.55$ G, which corresponds to the \si{} nuclei in the lattice site <111>, $a_k\approx1.5$ MHz, and which has the largest anisotropic SHI.\cite{Hale1969} There are strong clearly separated peaks only on the high field side of the hole. The peaks on the low field side are barely visible from the noise. This is closer to the pattern expected from the standard phonon modulated SHI.\cite{Pines1957} The same modulation effect was also seen in P doped silicon.\cite{Jarvinen2015} During hole burning with wide modulation a single FWIS step is not large enough for the spin packet to travel out of the pumping region, instead it is still inside the burning hole.
Assuming same probability $p$ for each FWIS flips, the probability
of the FWIS flips to remove the spin packet from the hole is reduced
to $\leq p^{2}$. A wider window leads to reduction of the FWIS
rate and, thus, increases the number of spin flips with higher SHI, mediated by the OE. This is clearly visible in the spectra with increasing modulation shown in fig. \ref{fig:OE_hole_mod}.

We think that in the present work with $^{75}$As both effects, the OE DNP and the frozen core softening are contributing to the observed hole shape.

\section{Spectral and spin diffusion}

Hole burning introduces a spatially inhomogeneous \si{} polarization
around the donors or depletion of a certain spin ensemble which was in resonance before burning (the frozen core softening). After burning, the hole and the shoulder peaks decay and broaden on a time scale ranging from tens of seconds to several days, depending on burning conditions and temperature of the sample. These effects are attributed to nuclear spin diffusion (NSD).\cite{Bloembergen1949} The electron spin spectral diffusion is orders of magnitude faster, and the \si{} nuclear relaxation is extremely slow at low temperatures of this work.\cite{Hayashi2009} We found that the decay rate of the SE holes is substantially slower than that of the OE holes. This can be explained
by the preferential polarization of the nuclei with large SHI specific for the SE DNP.

\begin{figure}
\includegraphics[width=0.8\columnwidth]{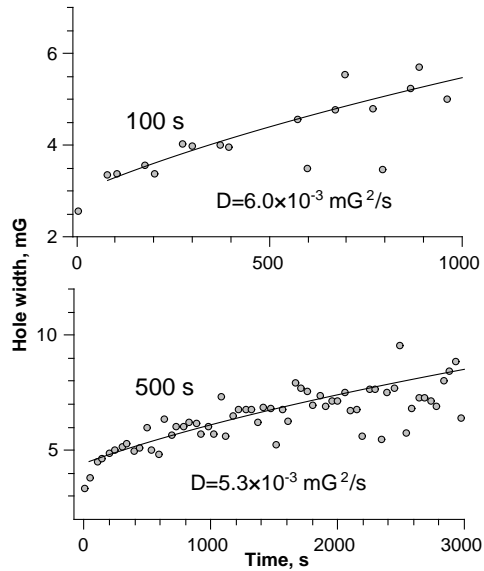}\caption{Widths of the OE holes after different hole burning times (marked
on the figures). The value of $D$ in the figures is extracted from
the fits shown with solid lines. The holes were burned with very small
(-73 dB) power at 450 mK.\label{fig:Widths-of-the-hole}}
\end{figure}

\begin{table}
\caption{Spectral diffusion coefficients extracted from the hole width for
different hole burning times.\label{tab:Spectral-diffusion-coefficients}}

\begin{tabular}{|c|c|c|c|}
\hline 
Burn time, s & $D$, $10^{-3}\frac{\text{G}^{2}}{\text{s}}$ & Burn time, s & $D$, $10^{-3}\frac{\text{G}^{2}}{\text{s}}$\tabularnewline
\hline 
\hline 
10 & 11.0 & 100 & 6.0\tabularnewline
\hline 
30 & 7.7 & 100 & 8.7\tabularnewline
\hline 
50 & 8.7 & 200 & 7.0\tabularnewline
\hline 
50 & 5.7 & 500 & 5.3\tabularnewline
\hline 
50 & 8.0 & 1000 & 10.0\tabularnewline
\hline 
\end{tabular}
\end{table}

\begin{figure}
\includegraphics[width=1\columnwidth]{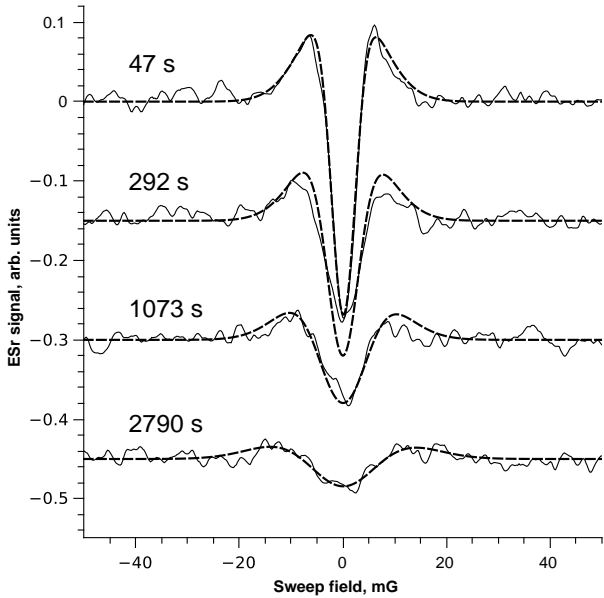}

\caption{Decay of OE hole after 500 s hole burn. The times marked on the curves
indicate the time delay from the end of the burn. The dashed solid
line is the calculated evolution of the hole from eq. (\ref{eq:spin diffusion}).
The signals are shifted 0.15 units from each other in the y-direction.\label{fig:OE_hole_decay_model}}
\end{figure}

\begin{figure}
\includegraphics[width=1\columnwidth]{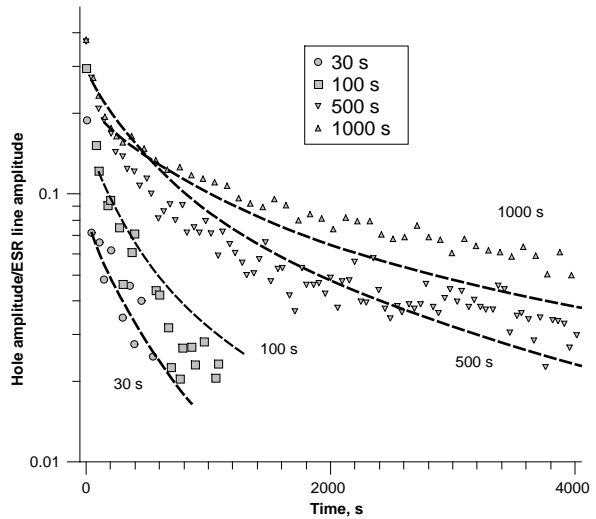}

\caption{Evolution of OE hole amplitudes after different hole burning times.
The holes were burned with very small (-73 dB) power at 450 mK. The
dashed lines are calculated decays of the holes with $D=8\times10^{-3}$
mG$^{2}$/s. \label{fig:OE-hole-evolution}}
\end{figure}

In this section we present results of the evolution of the holes created by the OE DNP. The dynamics of their behaviour depends on the pumping conditions, excitation power and pumping time. At high powers, the ESR transitions are strongly saturated and the instrumental effects related with the non-ideal spectrum of the mm-wave source start to be imprimted into the shape of the hole. Therefore, we performed the study of the hole relaxation after burning with the lowest power used for the excitation, varying the burning time only. In fig. \ref{fig:Widths-of-the-hole} we present an evolution of the width of the holes burnt for 100 and 500 sec. Fig. \ref{fig:OE_hole_decay_model} illustrates the change of the hole shape after 500 sec pumping time. In fig. \ref{fig:OE-hole-evolution} the log plot of the hole amplitudes as a function of time are plotted for different pumping times.

Full modeling of the spin diffusion would require quite complicated
calculations, which are out of scope of this study. Instead, in our analysis of the hole decay after the OE burning we employ a simple model based on one-dimensional spectral diffusion. In this model each spin packet in the spectrum is considered to follow a random walk process in which individual \si{}
spin flips move the spin packet to left or right.\cite{Wenckebach2016}
Macroscopically the motion of the spin packets can be then described
with the spectral diffusion equation:

\begin{equation}
\frac{\partial S(\omega,t)}{\partial t}=D\frac{\partial^2 S(\omega,t)}{\partial\omega^{2}}.\label{eq:spin diffusion}
\end{equation}

Here $S(\omega,t)$ is the ESR lineshape and $D$ is the spectral diffusion
constant. The diffusion constant for a one dimensional random walker
is given by
\begin{equation}
D=\frac{l^{2}}{2\delta t},\label{eq:spin diff coef}
\end{equation}
where $l$ is step length and $\delta t$ is time step between the
successive steps. The easiest way to estimate $D$ is to measure the
spreading of the hole as it decays. It is easy to show that the width at half minimum of a Gaussian hole, according to eq. (\ref{eq:spin diffusion}), evolves as
\begin{equation}
W_{h}=4\sqrt{D^{*}t\ln2+W_{0}},\label{eq:Hole width}
\end{equation}

where $t$ is time, $W_{0}$ is intrinsic linewidth and $D^{*}K=D$.
The constant $K$ takes into account the influence of the shoulders
near the OE hole. For a normal Gaussian hole $K=1$. The shoulders,
however, restrict the spreading of the hole and this leads
to smaller value of $D^{*}$ extracted from the measured $W_{h}$.
The value $K\approx3.1$ was estimated from the numerical solutions
of eq. (\ref{eq:spin diffusion}) using different values of $D$ and
eq. (\ref{eq:Hole-fit_function}) as the initial value. The examples
of fits are shown for two different burning times in fig. \ref{fig:Widths-of-the-hole}.
The $D$ values obtained from the fits for different burning times are shown in table (\ref{tab:Spectral-diffusion-coefficients}), with the mean value $D\approx(8\pm3)\times10^{-3}$ mG$^{2}$/s. The scatter of the spectral diffusion constant $D$ is fairly small considering that the hole burning time changes two orders of magnitude.

Next, we use eq. (\ref{eq:spin diffusion}) for analysis of the decay of the hole depth/amplitude after different length of burning times.
In all cases the hole decayed very quickly during the first $50 - 100$ s followed by a substantially slower part, as shown in fig. \ref{fig:OE-hole-evolution}. It was not easy to follow the hole shape during the fast evolving initial part. As well, we do not understand its origin. Therefore, we restricted our analysis to the slow part. Initial shape of the hole was taken $50 - 100$ s after burning, in the end of the fast part. Then, the eq. (\ref{eq:spin diffusion}) was then solved numerically. The simulated
decay of a hole after 500 s burning time together with the experimental data is shown in fig. \ref{fig:OE_hole_decay_model}. The fit of the first
hole, 47 s after the burning, is used as the initial value. The average
shape and spreading of the hole is predicted quite well with the spectral diffusion model.

In fig. \ref{fig:OE-hole-evolution} there are measurements of OE
hole decays after different burning times. The hole depth was determined
from the fits of the holes to eq. (\ref{eq:Hole-fit_function}). The
depth increases at longer burning times, and saturates at
around 500 s. Longer burns do not make the hole any deeper, only wider.
Also the decay of the hole gets slower for longer burning time. The
decay of the hole with short burn times is reasonably well described
by eq. (\ref{eq:spin diffusion}). With longer burning times, however,
the deviation is larger.

Spectral diffusion described above is caused by a spatial diffusion of the spin polarization, nuclear spin diffusion (NSD). Nuclear spin diffusion coefficient is given by \cite{Bloembergen1949}
\[
D_{s}=b^{2}W_{12},
\]
where $b$ is the average distance between two \si{} nuclei and $W_{12}$
is the flip-flop transition rate. For natural silicon (Wigner-Seitz
radius) $b \approx 0.47$ nm and using the literature value for $D_{s}\text{\ensuremath{\approx}}1.7\times10^{-14}\textrm{cm}^{2}/\text{s}$ in bulk silicon \cite{Hayashi2008}, we calculate $1/W_{12}\approx0.33$ s.  Using
radius of frozen core of about 8 nm, the time for the spin to diffuse
that distance is\cite{Hayashi2008}
\[
\Delta t=\frac{R^{2}}{4D_{s}}\approx9\text{ s}.
\]

These times can be compared with the time of the spin-packet jump $\delta t$ in the spectral diffusion model. For the smallest width of a hole of $\approx3$ mG, from eq. (\ref{eq:spin diff coef}) we estimate $\delta t\approx560$ s. This is substantially longer than the spin-diffusion flip-flop time, or even characteristic time of the spin diffusion over the distance of the frozen core. If the \si{} OE polarization would have taken place outside the frozen core or near its edge, the holes should disappear much faster than we observed. Thus our observation indicates that the diffusion is substantially slowed at the edge of the observation zone with $r_o\approx 5$ nm. Since this radius is smaller than the frozen core radius, this result does not look surprising and just supports the existence of the frozen core. However, the observed decays of the holes provide an evidence that the spin diffusion is not completely frozen even for the nuclei substantially closer to the donor than $r_o$. The remote peaks in the SE patterns corresponding to the DNP of the closest \si{} nuclei, do decay as well, although much slower than the OE holes. Measurement of the SE and OE DNP evolution may provide information on the spin dynamics inside the so-called frozen core. However the analysis of this spin dynamics is rather complicated due to several factors. For each donor the spin diffusion rate depends on the microscopic configuration of \si{} nuclei around it and is strongly anisotropic in magnetic field. The NSD depends on angles and distances between the \si{} spins and it can proceed with vastly different rates. In such situation, spin dynamics is somewhat similar to the systems with fractional diffusion channels, e.g. in NMR diffusion experiments in porous and fractal samples. \cite{Klemm1997,Klemm1999} Also, the frozen core is not at all a smooth sphere around the donor and the details of DNP effect inside the frozen core are not well understood. There are indications that at certain conditions spin diffusion can proceed even inside the barrier.\cite{Guichard2015,Wittmann2018} 
Rapid decay of OE holes immediately after burning may occur due to a fast spin diffusion for certain ensembles of donors with specific configurations ensuring easy flip-flops of the spins. The deviation between spectral and
spin diffusion is larger at longer burning times. This could
be caused by enhanced bulk polarization around the donors developed at long polarization time. 

\section{Conclusions}

We have studied the solid and Overhauser effect DNP of \si{} in $^{75}$As doped natural silicon in hole burning experiments.
SE DNP can be efficiently utilized to achieve high degree of polarization of
the \si{} nuclei closest  to the donor and having strong anisotropic SHI with the donor electron. On the contrary, the OE hole burning DNP experiments showed preferential flips of the remote \si{} spins with weakest SHI. This effect is not related to OE polarization and may be explained by the NSD and softening of the  frozen core at certain distance from the donor. This is caused by Rabi oscillations of the donor electron spin during saturation of the allowed ESR transitions required for the OE DNP. The polarization of spins with stronger SHI becomes visible only when the ESR excitation is widened with modulation. The decay of OE holes was analysed with a one-dimensional spectral diffusion model. The characteristic time of the spectral diffusion turns out to be substantially slower than the reported spin-diffusion time for the nuclei in the bulk of silicon crystal, far from the donors. This is a confirmation of the strong reduction of the spin diffusion rate inside the frozen core. Further studies of the details of the DNP process for the \si{} nuclei near donors may clarify the physics of the spin-dynamics in this system and find important applications in the field of quantum information processing.  

\bibliographystyle{aipnum4-1}
\bibliography{silicon}

\end{document}